\documentclass[preprint,aps,showpacs,preprintnumbers,amsmath,amssymb]{revtex4}%
\usepackage{graphicx}
\usepackage{amsmath}%
\usepackage{amsfonts}%
\usepackage{amssymb}
\begin{document}

\title{Polymers with self-avoiding interaction in random medium: \\a localization-delocalization transition }
\author{Yadin Y. Goldschmidt$^{1}$ and Yohannes Shiferaw$^{1}$,$^{2}$}
\affiliation{$^{1}$Department of Physics and Astronomy, University of Pittsburgh,
Pittsburgh PA 15260}
\affiliation{$^{2}$Physics Department, Northeastern University, Boston, MA 02115}
\date{\today }

\begin{abstract}
In this paper we investigate the problem of a long self-avoiding polymer chain
immersed in a random medium. We find that in the limit of a very long chain
and when the self-avoiding interaction is weak, the conformation of the chain
consists of many ``blobs'' with connecting segments. The blobs are sections of
the molecule curled up in regions of low potential in the case of a Gaussian
distributed random potential or in regions of relatively low density of
obstacles in the case of randomly distributed hard obstacles. We find that as
the strength of the self-avoiding interaction is increased the chain undergoes
a delocalization transition in the sense that the appropriate free energy per
monomer is no longer negative. The chain is then no longer bound to a
particular location in the medium but can easily wander around under the
influence of a small perturbation. For a localized chain we estimate
quantitatively the expected number of monomers in the ``blobs'' and in the
connecting segments.

\end{abstract}%

%TCIMACRO{\TeXButton{pacs}{\pacs{36.20.Ey, 05.40-a, 75.10.Nr, 64.60.Cn}}}%
%BeginExpansion
%EndExpansion%
%TCIMACRO{\TeXButton{maketitle}{\maketitle}}%
%BeginExpansion
\maketitle
%EndExpansion

\section{Introduction}

The properties of polymer molecules trapped in a random environment like a gel
network \cite{asher} or inside porous materials or membranes
\cite{cannell,rondelez,bishop} is of considerable interest. On the theoretical
side many papers were written on the subject in an effort to understand the
basic properties of the model, like the radius of gyration (or end-to-end
distance) of a single molecule immersed in the random medium [5-14]. In
several papers an ideal (Gaussian) chain has been used, which corresponds
approximately to the experimental situation at the so called $\Theta
$-temperature when the solvent effectively screens the self-avoiding
interaction of the chain. Even in that case the properties of a single
molecule in a random potential are not simple. First, the properties depend on
the type of randomness introduced- like a Gaussian random potential versus a
sea of hard obstacles. Second, it may depend on whether the random potential
is annealed or quenched, i.e. if the impurities are free to move or fixed.
Additionally, different results apply if the chain is anchored or free to
move. It turns out that for quenched randomness and a chain that is free to
move, the size of the total volume of the medium is important for determining
the chain size. For example, in the case of a random potential with a Gaussian
distribution, the chain size saturates as a function of its length $L$ (number
of monomers) for a very long chain, in contrast to a dependence like
$R_{g}=aL^{1/2}$ (where $R_{g}$ is the radius of gyration and $a$ is the
monomer size) in the absence of a random potential . A logarithmic dependence
of the chain size on the volume of the medium was argued in reference
\cite{cates} using qualitative arguments. An analytic derivation of the
logarithmic dependence was derived in reference \cite{gold} using the replica
method and the variational approximation, with a replica-symmetry-breaking
solution. In reference \cite{shifgold} the result was justified from a mapping
of the polymer problem to the problem of the localization of a quantum
particle in a random potential. In reference \cite{goldshif} results were
obtained for a polymer immersed in a sea of random obstacles, and various
behaviors were identified as a function of the system size. Again the use of
localization theory and extreme value statistics were instrumental to the derivation.

In some of the papers \cite{BM,nattermann,thirumalai} there was an attempt to
include the effects of the self-avoiding interaction of the polymer. These
attempts were far from complete. For example in Ref. \cite{nattermann} it was
assumed that the conformation of the polymer consists of one spherical blob
and it was argued that a quenched random potential is irrelevant for a very
long chain. In Ref. \cite{thirumalai} analytical results were obtained for
annealed disorder, and simulations were performed for strictly self-avoiding
walks. Ref. \cite{BM} presents numerical evidence for a size transition of the
polymer as a function of the relative strength of the disorder and the
self-avoiding interaction. The simulations were carried out for a random
distribution of hard obstacles with a concentration exceeding the percolation
threshold. It is the aim of the present paper to shed more light on this
important problem. We will make use mainly of Flory-type arguments, and
consider both the case of a Gaussian random potential and the case of randomly
placed obstacles. A polymer with self avoiding interaction can not be mapped
into a quantum particle at a finite temperature in a simple manner, because
for a quantum particle there no impediment to return at a later time (or
Trotter time) to a position it visited previously.

An important point to keep in mind is the strength of the excluded volume
interaction. If one consider a strictly self-avoiding walk on a lattice (SAW)
corresponding to a non-self-intersecting chain, then the strength of the
Edwards parameter $v$ \cite{doi} is fixed at $O(1)\times a^{3}$ where $a$ is
the step size (or monomer size) and depends only on the type of lattice. On
the other hand one can consider a Domb-Joyce model \cite{domb} where there is
a finite penalty for self overlapping of polymer segments, and then the
strength of $v$ can be varied substantially and reduced continuously to zero.
The interplay between the strength of the self avoiding interaction and the
strength of the disorder can then be investigated to a larger extent.
Experimentally the Edwards parameter is given approximately \cite{degennes} by
$v=a^{3}(1-2\chi)$, where $\chi$ is the Flory interaction parameter, which
depends on the chemical properties of the polymer and the solvent, and on the
temperature (and pressure). It takes the value 1/2 at the $\Theta$-point. The
case $\chi=0$ corresponds to a solvent that is very similar to the monomer. In
general good solvents have low $\chi$ whereas poor solvents have high $\chi$
resulting in $v$ being negative. In the following we will restrict ourselves
to the case of positive $v$, which leads to the more interesting and
non-trivial results.

Before we proceed to investigate the effects of the self-avoiding interaction,
let us summarize in more detail the known results for a free Gaussian chain in
a quenched random potential with a Gaussian distribution with variance $g$,
and for a free Gaussian chain in a sea of fixed random obstacles with average
concentration $x$ per site.

For an uncorrelated Gaussian random potential, it was argued using qualitative
arguments in Refs. \cite{cates,nattermann} that a very long Gaussian chain,
that is free to move in the medium, will typically curl up in some small
region of low average potential. The polymer chain is said to be localized,
and for long chains the end-to-end distance ($R_{F}$) becomes independent of
chain length $L$ (number of monomers) and scales like
\begin{equation}
R_{F}\propto(g\ln\mathcal{V})^{-1/(4-d)}, \label{rfg}%
\end{equation}
where $\mathcal{V}$ is the volume of the system, $d$ is the number of spatial
dimensions, and the random potential satisfies $\langle U({\mathbf{x}%
})U(\mathbf{x}^{\prime})\rangle=g\delta(\mathbf{x}-\mathbf{x}^{\prime})$. The
binding energy per monomer of the chain is given approximately by
\begin{equation}
U_{bind}/L\sim(g\ln\mathcal{V})^{2/(4-d)}. \label{gsg}%
\end{equation}
These results were also obtained by the replica method in Ref. \cite{gold},
and rederived using a mapping to a quantum particle's localization in Ref.
\cite{shifgold}. For very short chains the end-to-end distance scales
diffusively ($R_{F}^{2}\sim L$), and it saturates at the $R_{F}$ value quoted
above for large $L$. Notice, that in the infinite volume limit, the chain
completely collapses. This results from the fact that the depth of the
potential is unbounded from below, and the chain is always able to find with
reasonable probability a deep enough narrow potential well to occupy,
overcoming its tendency to swell due to the entropy of confinement. The
collapse of the chain in the infinite volume limit agrees with the results for
a chain in an annealed potential, since the ability of a free chain to scan
all space for a favorable environment is equivalent to the random potential
adapting itself to the chain configuration.

In contrast, a tethered chain that is anchored at one end behaves very
differently. Such a chain has a tadpole structure. The end of its tail is
anchored and its head is a curled coil situated at a deep potential well. Such
a chain has an end-to-end distance of order
\begin{equation}
R_{T}\sim\frac{g^{1/(4-d)}L}{(\ln L)^{1-1/(4-d)}},
\end{equation}
which is ballistic in three dimensions, since the chain has to look far to
find a low potential. Here there is no dependence on the system's volume and
the quenched and annealed cases are very different from each other even in the
infinite volume limit.

In reference \cite{goldshif} we considered the effect of randomly distributed
obstacles on the behavior of a free Gaussian chain. We have shown that in the
presence of infinitely strong obstacles, that totally exclude the chain from
visiting sites occupied by obstacles, with average concentration $x$ per site,
there are three possible behaviors of the end-to-end distance depending on the
system's volume $\mathcal{V}$. For $2<d<4$, if the volume of the system is
smaller than $\mathcal{V}_{1}\simeq\exp(c_{1}x^{-(d-2)/2})$, then the polymer
is localized, as in the case of a Gaussian random potential, and $R_{I}%
\sim\left(  x\ln\mathcal{V}\right)  ^{-1/(4-d)}$. For $\mathcal{V}%
_{1}<\mathcal{V}<\mathcal{V}_{2}$, where $\mathcal{V}_{2}\simeq\exp\left(
c_{2}x^{2/(d+2)}L^{d/(d+2)}\right)  $, the polymer size is given by
$R_{II}\sim\left(  x/\ln\mathcal{V}\right)  ^{-1/d}$ ($c_{1}$ and $c_{2}$ are
constants of order unity). Finally, for $\mathcal{V>V}_{2}$, the polymer
behaves the same way as for an annealed potential, and $R_{III}\sim\left(
L/x\right)  ^{1/(d+2)}$. Results are given to leading order in $x$ for small
$x$. These results are valid only when the average volume fraction of the
obstacles ($x$) is smaller than the percolation threshold. When $x$ is bigger
than the percolation threshold we expect that the system breaks into
independent domains whose volume is independent of, and generally much smaller
than the total volume of the system.

We end the introduction by revisiting the meaning of localization for polymers
in a random medium. Although some authors connect the compact size of the
chain when $L\rightarrow\infty$ with the notion of localization, this is
actually not so. The compact size should be viewed as a separate feature from
the notion of localization. It is rather the binding energy of a chain
$U_{bind}$ that has to exceed the translational entropy $\ln\mathcal{V}$. From
Eq. (\ref{gsg}) this amounts to the condition
\begin{equation}
\ln\mathcal{V}<L(g\ln\mathcal{V})^{2/(4-d)},
\end{equation}
which holds for any $2\leq d<4$ when $L$ is large enough (for $2<d<4$ and any
fixed $L$, the condition can be satisfied for large enough $\mathcal{V}$) .
This condition assures that the polymer will stay confined at a given location
and will not, under a some small perturbation move to a different location.
Thus repeating an experiment or a simulation with the same fixed realization
of the disorder, but with different initial conditions, will result in finding
the polymer situated at the same region of the sample as in a previous
experiment, provided of course one waits enough time (which can be enormous)
for the system to reach equilibrium. We observe that this condition is
satisfied for large enough $L$ provided the binding energy per monomer is
positive. Another interpretation of the inequality given above in the context
of equilibrium statistical mechanics is that the partition sum is dominated by
the term involving the ground state as opposed to the contribution of the
multitude of positive energy extended states. The contribution of these states
is proportional to the volume of the system and thus the inequality above
results from the condition
\begin{equation}
\exp(-LE_{0})>\mathcal{V}.
\end{equation}
What we will see in the following sections is that in the presence of a
self-avoiding interaction, a localization-delocalization transition occurs
when varying the strength of the the self-avoiding interaction for a fixed
amount of disorder or alternatively upon varying the strength of the disorder
for a fixed value of the self-avoiding interaction.

\section{Chains with self-avoiding interaction-the case of a Gaussian random potential}

We now study the case of a self-avoiding chain. In this section we consider
the case of a random potential with a Gaussian distribution, and postpone to
the next section the investigation of the case of a random distribution of
hard obstacles. For simplicity, the discussion in the rest of the paper will
be limited to three spatial dimensions ($d=3$).

We start with a quenched random potential with a Gaussian distribution with
variance $g$ at each lattice site. The lattice constant, being equal to the
monomer size, is $a$ and we will measure all lengths in units of $a$. We will
also measure all energies in units of $kT$ which is equivalent to setting
$kT=1$. We first recall the beautiful argument given by Cates and Ball
\cite{cates} for an ideal chain with no self-avoiding interaction. If we
coarse-grain the system and divide it into regions of volume $\sim R^{3}$, the
(average) potential in such a region has a Gaussian distribution with variance
$g/R^{3}$. Thus the distribution of the potential strength $V$ for such
coarse-grained volumes is given by
\begin{equation}
P(V)\sim\exp\left(  \frac{-R^{3}V^{2}}{g}\right)  .
\end{equation}

Here and in what follows we omit constants of order unity. If the volume of
the system is $\mathcal{V\ }(\gg R^{3})$, the lowest (negative) value $U$ of
the coarse-grained potential that is likely to be found in such a volume is
given by%
\begin{equation}
\int_{-\infty}^{U}P(V)dV=R^{3}/\mathcal{V},
\end{equation}
which yields the estimate%
\begin{equation}
R^{3}U^{2}/g\sim\log(\mathcal{V}),
\end{equation}
or%
\begin{equation}
U=-\sqrt{\frac{g\log(\mathcal{V})}{R^{3}}}. \label{minp}%
\end{equation}
This is the energy per each monomer which resides in this low potential
region. Thus if our ideal polymer chain is assumed to have a spherical shape
of diameter $\sim R$, it will curl itself in the region of the lowest
potential given by Eq.(\ref{minp}) above. To determine the volume $R^{3}$
occupied by the chain we recall that we have to pay in confinement entropy for
compacting the chain below its natural size. Thus we have to balance the
confining entropy versus the gain in potential energy from the random
potential. Thus the free energy is given by%
\begin{equation}
F\approx\frac{L}{R^{2}}-L\sqrt{\frac{g\log(\mathcal{V})}{R^{3}}}.
\label{freeenergy}%
\end{equation}
Notice that an unimportant constant, resulting from the entropy of a free
chain, has been dropped ($S_{0}=$ $L\log(z)$ on a lattice with $z$ the number
of nearest neighbors, which contributes $-S_{0}$ to the free energy. This term
does not depend on $R$ and does not contribute to the binding energy). The
optimal size $R_{m}$ is found by minimizing $F$ and is found to be%
\begin{equation}
R_{m}\sim\frac{1}{g\log(\mathcal{V})}\equiv\frac{1}{G(\mathcal{V})}.
\end{equation}
where we defined the volume dependent disorder strength by $G(\mathcal{V}%
)=g\ln(\mathcal{V})$. Substituting this result in $F$ we obtain%
\begin{equation}
F_{m}=-\frac{L}{3R_{m}^{2}}\approx-G(\mathcal{V})^{2}L.
\end{equation}
We see that $-F_{m}/L$ is the binding energy per monomer, and it is strictly
positive, so the polymer is localized. In what follows we will assume that $g$
is small enough so that $G\ll1$ for the given system volume, hence $R_{m}\gg
1$, and the chain is not totally collapsed unless $\mathcal{V}\rightarrow
\infty$.

We now add a self avoiding interaction and assume first that it is small, i.e.
$v<<g$, or at least $v<g$. If the chain is still localized in the same well,
which we will see momentarily not to hold when $L$ is large, then%
\begin{equation}
F=\frac{L}{R_{m}^{2}}-L\sqrt{\frac{g\log(\mathcal{V})}{R_{m}^{3}}%
}+\frac{vL^{2}}{R_{m}^{3}}\approx-G(\mathcal{V})^{2}L+vG(\mathcal{V})^{3}%
L^{2}. \label{fsa}%
\end{equation}
Here, beside assuming that $v$ is small we assume for the moment that $L$ is
not too big so the last term in the free energy, resulting from the
self-avoiding interaction, is small enough so one does not have to take into
account the change in $R_{m}$ due to the presence of $v$. If we plot $F$ vs.
$L$, we see that it is lowest when
\begin{equation}
L=L_{m}=\frac{1}{2vG(\mathcal{V})}.
\end{equation}
Thus if $vg\ll1$, we have $L_{m}\gg1$, and
\begin{equation}
F_{m}=-\frac{G(\mathcal{V})}{4v}.
\end{equation}
For $L=2L_{m}$ the free energy vanishes and for larger $L$ it eventually
increases fast like $L^{2}$. We can now verify that if $L$ does not exceed
$L_{m}$ then the approximation used above, assuming that $R_{m}$ does not
change appreciably from its ideal chain value, is justified. If we
differentiate the above $F$ in Eq.\ (\ref{fsa}) with respect to $R_{m}$ we
find%
\begin{equation}
R_{m}\sim\frac{1}{G(\mathcal{V})}\left(  1+\frac{vL}{R_{m}}\right)
^{2}\approx\frac{1}{G(\mathcal{V})}\left(  1+vG(\mathcal{V})L\right)  ^{2},
\end{equation}
again omitting constants of order unity. Thus the correction $vGL$ evaluated
at $L=L_{m}$ is of order unity, and we can still use the value $R_{m}\sim1/G$.

For larger $L$ the approximation seems to break down, but fortunately what
happens is that since the free energy increases when $L$ exceeds $L_{m}$, it
is energetically favorable for part of the chain to jump into a distant well.
Even though there is a cost for the polymer segment between the wells one
still gains in the overall free energy from the binding energy in the wells.
Thus the picture that emerges is that as $L$ increases, the chain divides
itself into separate blobs with connecting segments. In each blob the number
of monomer does not exceed $L_{m}$, which is the optimal value for that well.
The idea is depicted in Figure 1.%

%TCIMACRO{\FRAME{ftbpFU}{4.779in}{3.1825in}{0pt}{\Qcb{Illustration of how the
%low free energy conformations evolve with increasing chain length. The chain
%length increases with the direction of the arrows.}}{\Qlb{fig1}}%
%{blobs.eps}{\special{ language "Scientific Word";  type "GRAPHIC";
%maintain-aspect-ratio TRUE;  display "USEDEF";  valid_file "F";
%width 4.779in;  height 3.1825in;  depth 0pt;  original-width 9.5008in;
%original-height 6.3079in;  cropleft "0";  croptop "1";  cropright "1";
%cropbottom "0";  filename 'blobs.eps';file-properties "XNPEU";}}}%
%BeginExpansion
\begin{figure}
[ptb]
\begin{center}
\includegraphics[
height=3.1825in,
width=4.779in
]%
{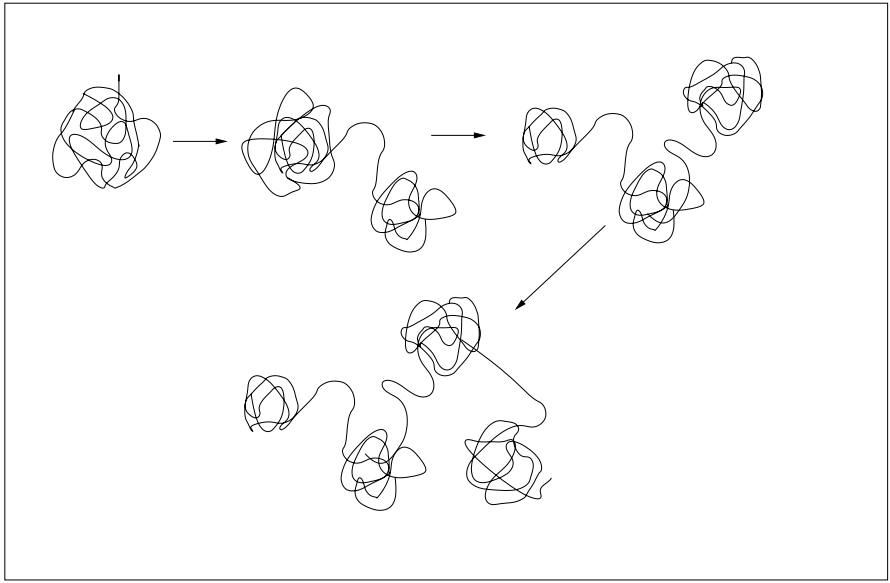}%
\caption{Illustration of how the low free energy conformations evolve with
increasing chain length. The chain length increases with the direction of the
arrows.}%
\label{fig1}%
\end{center}
\end{figure}
%EndExpansion

To be more specific we will now construct a model for the free energy of the
chain. The first blob will be located in the deepest minimum in the total
volume, whose depth is roughly given by $-G(\mathcal{V})^{2}$ per monomer,
where $\mathcal{V}$ is the total volume of the system. Subsequent blobs will
reside in the most favorable well within a range $Y$, which has to be taken
self-consistently as the length of the jump. Thus within a range $Y$ the chain
is likely to find a potential minimum of depth $-G(Y^{3})^{2}$. The farther
you jump, the deeper well you are likely to find. Thus assuming for simplicity
that the jumps are roughly of equal size, and there are $K$ blobs in addition
to the initial blob, the free energy of the chain will be given roughly by%
\begin{equation}
F(w,m,Y;L)\approx-\frac{G(\mathcal{V})}{4v}+K\left(  \frac{Y^{2}}{m}%
+\frac{m}{Y^{2}}+\frac{vm^{2}}{Y^{3}}-wG(Y^{3})^{2}+vw^{2}G(Y^{3})^{3}\right)
,
\end{equation}
with%
\begin{equation}
K=\frac{L-L_{0}}{w+m},\;L_{0}=\frac{1}{2vG(\mathcal{V})},\;G(Y^{3})=3g\ln(Y).
\end{equation}
We defined $w$ to be the number of monomers in each blob, and $m$ to be the
number of monomers in each connecting segment. The term $Y^{2}/m$ results from
the ``stretching'' entropy of the segment and $m/Y^{2}$ from the confinement
entropy. The term $vm^{2}/Y^{3}$ represents the self-avoiding interaction for
the connecting segments. $L_{0}$ is the number of monomers in the initial blob
whose free energy was taken care of separately. It is evident that when $L$ is
very large we can neglect the free energy of the first blob and also take
$K\approx L/(w+m)$. Thus we find for the free energy per monomer%
\begin{equation}
f(w,m,Y)\equiv\frac{F(w,m,Y;L)}{L}\approx\frac{1}{w+m}\left(  \frac{Y^{2}}%
{m}+\frac{m}{Y^{2}}+\frac{vm^{2}}{Y^{3}}-wG(Y^{3})^{2}+vw^{2}G(Y^{3}%
)^{3}\right)  .\label{fepm}%
\end{equation}
This function has to be minimized with respect to $w$, $m$ and $Y$ to find the
parameters giving rise to its lowest value. For the connecting pieces of the
chain we did not include a contribution from the random potential since it is
expected to average out to zero for these parts. To gain some feeling into the
behavior of this function and the values of the parameters which minimize it,
we display in Table I the value of the parameters and free energy per monomer
for $g=0.05$ and various values of $v$, as obtained from a minimization
procedure. The delocalization transition is the point where $f$ changes sign
from negative to positive, as discussed earlier. Actually, to be more precise,
the delocalization transition occurs when $f=-(\ln\mathcal{V})/L$ for finite
$L$, when the translational entropy starts to exceed the binding energy. In
the limit of large $L$ we can say that the transition is at $f=0$. We observe
that the delocalization transition occurs at $v=0.0478$ which is close to the
value of $\ g$. We also observe that $m\ll w$ for $v\ll g$ and $m\gg w$ near
the transition. Also for small $v$, $m\sim Y$, whereas near the transition
$m\sim Y^{2}$. If we compare the value of $w$ from Table I with the value of
$L_{m}=1/(2vG(Y^{3}))$ we find that $w$ is smaller than $L_{m}$ in the entire
range. The ratio $w/L_{m}$ varies from $\sim0.4$ to 1 as $v$ changes from
$10^{-5}$ to 0.048. Thus the assumption we have made previously concerning
this ratio is justified \textit{a posteriori}.

\begin{table}[ptb]%
\begin{tabular}
[c]{|l|l|l|l|l|}\hline
v & Y & m & w & f\\\hline
0.00001 & 2206 & 2413 & 16178 & -0.835249\\\hline
0.0001 & 346 & 534 & 2580 & -0.421023\\\hline
0.001 & 60 & 148 & 461 & -0.164673\\\hline
0.01 & 13.7 & 66 & 97 & -0.0370883\\\hline
0.02 & 11.1 & 69.2 & 60.8 & -0.0156444\\\hline
0.03 & 12.8 & 105.5 & 41.9 & -0.0056033\\\hline
0.04 & 22 & 302.6 & 26.8 & -0.0009992\\\hline
0.045 & 35.6 & 720.1 & 20.7 & -0.0001776\\\hline
0.0478 & 69.4 & 2342.4 & 16.5 & 0\\\hline
0.048 & 77 & 2800 & 16 & 0.00000725\\\hline
\end{tabular}
\caption{Parameters and free energy for the case g=0.05}%
\label{table1}%
\end{table}

Luckily it was possible to solve the minimization equations analytically
almost entirely in both the limits $v\ll g$, and near the transition when
$f\approx0$. Details of the solutions are given in the Appendix. Here we only
display the results:

A. The case $v\ll g$.

The parameters are given by%
\begin{align}
Y  &  =\frac{2}{v}(\ln Y-1)^{-1/2}(\ln Y+3)^{-3/2},\label{yvs}\\
m  &  =\frac{2}{3vg\ln Y}(\ln Y-1)^{-1}(\ln Y+3)^{-1},\label{mvs}\\
w  &  =\frac{2}{3vg\ln Y}(\ln Y+3)^{-1}\label{wvs}\\
f  &  =-9g^{2}(\ln Y)^{2}(\ln Y-1)(\ln Y+3)^{-1}. \label{fvs}%
\end{align}
The first equation can be easily solved numerically for $Y$ for a given value
of $v$ and the result substituted in the other equations. Very good agreement
is achieved with Table I for small values of $v$.

B. Solution near the delocalization transition.

Let us define the parameter%
\begin{equation}
\kappa=\frac{4v}{3g}. \label{kappa}%
\end{equation}
In terms of this parameter we have%
\begin{align}
Y  &  =\frac{1}{v\kappa}\left(  \frac{1+\sqrt{1+\kappa^{2}}}{\kappa}\right)
^{2},\label{ytr}\\
m  &  =\frac{1}{v^{2}\kappa^{2}}\left(  \frac{1+\sqrt{1+\kappa^{2}}}{\kappa
}\right)  ^{3},\label{mtr}\\
w  &  =\frac{\kappa}{8v^{2}\ln Y}=\frac{1}{2vG(Y^{3})}. \label{wtr}%
\end{align}
The transition point is obtained by solving the equation%
\begin{equation}
\frac{1+\sqrt{1+\kappa^{2}}}{\kappa}+\frac{\kappa}{1+\sqrt{1+\kappa^{2}}%
}+\frac{1}{\kappa}\left(  1-2\ln\left(  \frac{2(1+\sqrt{1+\kappa^{2}})}%
{\sqrt{3g}\kappa^{2}}\right)  \right)  =0. \label{feqz}%
\end{equation}
Once the solution $\kappa_{c}$ is determined for a given $g$, then the
transition point $v_{c}$ is determined from $v_{c}=3g\kappa_{c}/4$. For $g$ in
the range 0.01-0.2 we find that $\kappa_{c}$ is a number of order unity
(varies from 1.7 to 0.96 as $g$ changes in that range). This means that
$v_{c}$ is quite close to $g$. Once $v_{c}$ is known, all the parameters $Y,m$
and $w$ at the transition are determined by the solution above. For $g=0.05$
we get $\kappa_{c}=1.2744$, and thus $v_{c}=0.0478$ in excellent agreement
with the minimization results from Table I.

For $v>v_{c}$ the chain is delocalized. The above expression for the free
energy may no longer be accurate, but the general picture is clear. There will
be very few monomers in the low regions of the potential, and the chain will
behave very much like an ordinary chain with a self-avoiding interaction in
the absence of a random potential. Any little perturbation can cause the chain
to move to a different location in the medium (see Fig. 2).%

%TCIMACRO{\FRAME{ftbpFU}{4.5316in}{3.3036in}{0pt}{\Qcb{A typical chain
%conformation when $v\gg g$. The dark regions are regions of low average
%potential. Only short segments of the chain are situated in these regions.}%
%}{\Qlb{fig2}}{avoiding.eps}{\special{ language "Scientific Word";
%type "GRAPHIC";  maintain-aspect-ratio TRUE;  display "USEDEF";
%valid_file "F";  width 4.5316in;  height 3.3036in;  depth 0pt;
%original-width 9.8779in;  original-height 7.1857in;  cropleft "0";
%croptop "1";  cropright "1";  cropbottom "0";
%filename 'avoiding.eps';file-properties "XNPEU";}} }%
%BeginExpansion
\begin{figure}
[ptb]
\begin{center}
\includegraphics[
height=3.3036in,
width=4.5316in
]%
{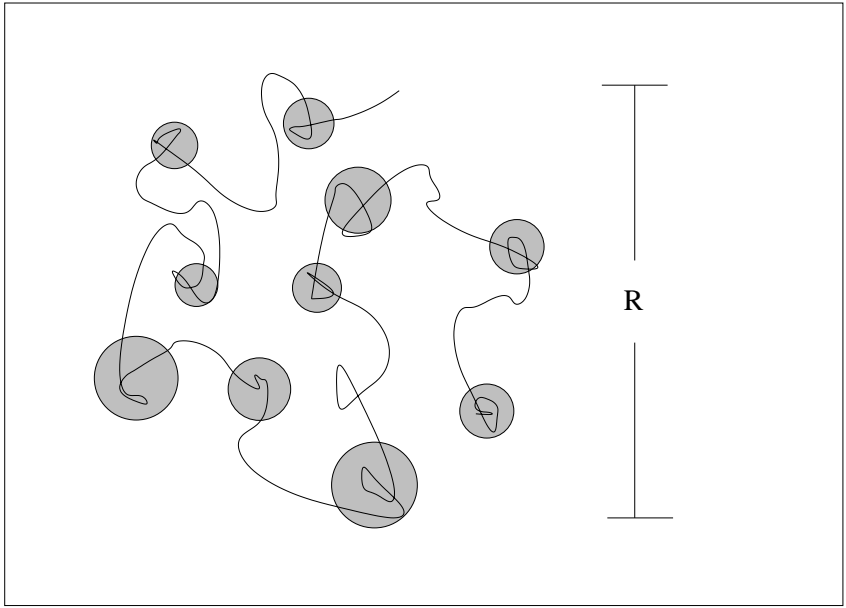}%
\caption{A typical chain conformation when $v\gg g$. The dark regions are
regions of low average potential. Only short segments of the chain are
situated in these regions.}%
\label{fig2}%
\end{center}
\end{figure}
%EndExpansion

For small values of $v$ when the chain is localized, we still expect its size
to grow like that of a self-avoiding walk. Thus we expect roughly
\begin{equation}
R_{g}\sim YK^{0.6},
\end{equation}
since $Y$ is the step size, and $K$ is the number of steps. But since
$K\approx L/(m+w)$ for large $L$, we find%
\begin{equation}
R_{g}\sim\frac{Y}{(m+w)^{0.6}}L^{0.6}.
\end{equation}
Thus the chain behaves as a self avoiding walk with an effective step size per
monomer given by
\begin{equation}
\frac{a_{eff}}{a}=\frac{Y}{(m+w)^{0.6}}.
\end{equation}
From Table I it becomes clear that the effective monomer size changes from a
value of $6$ for $v=0.00001$ to a value of $\sim0.66$ at the transition. The
reason for the large value of the effective monomer size at very small $v$ is
that the chains makes long jumps to take advantage of deep wells of the random
potential, very much like anchored chains in a random potential that make
sub-ballistic jumps \cite{cates}. This is also the reason why the number of
monomers $w$ in each well is somewhat less than $L_{m}$ , since a sufficient
number of monomers need to be used for the connecting segments. For $v\gg v_{c}$
in the delocalized phase we expect the effective monomer size to be about 1,
since the chain behaves almost like an ordinary self-avoiding chain with the
random potential not playing any significant role. Thus the chain is expected
to have its smallest size in the vicinity of the transition.

It is important to notice that the discussion above is in the limit for very
large $L$. If $L\lesssim1/(2vG(\mathcal{V}))$, then in the localized phase
when $v\ll v_{c}$ the polymer will be confined to a single well and will
appear compact, even though it will not remain so for large $L$. This may
explain why in simulations that were done typically with $L\leq320$ \cite{BM}
the delocalization transition appeared as a transition from a compact to a
non-compact state of the polymer.

We should also note that for an annealed random potential there is a
transition from a collapsed state into an ordinary self-avoiding chain as $v$
increases through the point $v=g$ \cite{nattermann}. This is because the
annealed free energy reads%
\begin{equation}
F(R)=\frac{L}{R^{2}}-\frac{gL^{2}}{R^{3}}+\frac{vL^{2}}{R^{3}}+\frac{R^{2}}%
{L}.
\end{equation}
For $v<g$ the fourth term is negligible and the free energy is lowest when
$R\rightarrow0$ (when $L$ is large). For $v>g$ the first term is negligible,
and the radius of gyration grows like%
\begin{equation}
R\sim(v-g)^{1/3}L^{3/5}.
\end{equation}

\section{The case of hard obstacles}

The case of hard obstacles was investigated recently by us in Ref.
\cite{goldshif}. As already discussed in the introduction, three different
behaviors were identified as a function of the system's volume. Region I is
defined when the system's volume $\mathcal{V}<\mathcal{V}_{1}\sim\exp
(c_{1}/\sqrt{x})$, where $c_{1}$ is a constant of order unity. Here $0<x<1$ is
the average concentration of obstacles per site (total number of obstacles
divided by total number of sites). We also assume that $x$ is less than the
percolation threshold ($x_{c}=0.3116$ for a cubic lattice), so sites occupied
by obstacles don't percolate. In Region I we recall \cite{goldshif} that in
the absence of a self-avoiding interaction, the free energy per site for a
chain situated in a spherical region of volume $R^{3}$ in three dimensions is
given by
\begin{equation}
F_{I}/L=-\ln(z)+1/R^{2}+\hat{x}%
\end{equation}
where $\hat{x}$ is the actual concentration of obstacles in that region, whose
minimal expected value in a system of total volume $\mathcal{V}$ is
\begin{equation}
\hat{x}_{m}\simeq x-\sqrt{\frac{x\ln\mathcal{V}}{R^{3}}}.
\end{equation}
The binding energy per monomer inside the blob resulting from the lower
concentration of obstacles in this region is given by $x-\hat{x}$, since it is
equal the entropy gain from a lower concentration as compared to the average
(background) concentration $x$. The chain is ``sucked'' towards regions with
low concentration of obstacles since it can maximize its entropy there, and
these regions of space act like the negative potential regions of the Gaussian
random potential. Thus in order for the free energy per monomer to reflect
correctly the binding energy of the chain inside the blob, both the constant
$x$ of the background and the constant term $-\ln(z)$, which is always there
regardless of the chain's position, have to be subtracted. The relevant free
energy per monomer situated in the blob (which is equal to minus the binding
energy) is given by
\begin{equation}
f_{I}=\frac{1}{R^{2}}-\sqrt{\frac{x\ln\mathcal{V}}{R^{3}}}.
\end{equation}
This result coincides with Eq. (\ref{freeenergy}) upon the substitution
$g\rightarrow x$. Thus all the results of the previous section carry on to
Region I with this simple substitution .

Therefore we are going to discuss the situation when the system's volume is
greater than $\mathcal{V}_{1}$ (Region II). In this case the many blob picture
still holds, where the blobs are now situated in regions free of obstacles
(with $\hat{x}=0$ ) whose size is determined again by the distance $Y$ of the
jump which is also assumed to satisfy $Y^{3}\gtrsim\mathcal{V}_{1}$ ( an
assumption which will be justified \textit{a posteriori}). In this case the
farther the jump, it is more likely for the chain to find a larger space empty
of obstacles, which will reduce further its confinement entropy $w/R^{2}$ and
also the self-avoiding energy ( but there is a cost resulting from the
connecting segments and the constraint of the total length being fixed). The
blob size is given by $R_{mII}=1/G_{o}$ (the largest expected empty region in
a volume $Y^{3}$ ) with \cite{goldshif}%
\begin{equation}
G_{o}(Y^{3})\approx\left(  \frac{x}{3\ln Y}\right)  ^{1/3}.
\end{equation}
Thus the free energy per monomer of a chain consisting of a number of blobs
with connecting parts, each of length $Y$,  is given by%
\begin{equation}
f(w,m,Y)\approx\frac{1}{w+m}\left(  \frac{Y^{2}}{m}+\frac{m}{Y^{2}%
}+\frac{vm^{2}}{Y^{3}}-w(x-G_{o}(Y^{3})^{2})+vw^{2}G_{o}(Y^{3})^{3}\right)  ,
\end{equation}
where again the constant $x$, which is independent of $m$, $w$ and $Y$, has been
subtracted. This assures that the delocalization transition again occurs at
$f=0$ and not at $f=x$. The results of a numerical minimization of this free
energy is displayed in Table II for the case of $x=0.1$.

\begin{table}[ptb]%
\begin{tabular}
[c]{|l|l|l|l|l|}\hline
v & Y & m & w & f\\\hline
0.00001 & 552 & 2206 & 72092 & -0.062\\\hline
0.0001 & 127 & 565 & 9135 & -0.051\\\hline
0.001 & 38 & 206 & 1241 & -0.034\\\hline
0.01 & 23 & 230 & 207 & -0.0077\\\hline
0.02 & 32 & 536 & 137 & -0.0019\\\hline
0.03 & 59 & 1737 & 120 & -0.00035\\\hline
0.04 & 194 & 13915 & 131 & -0.0000099\\\hline
0.041 & 248 & 21327 & 134 & -0.0000023\\\hline
0.042 & 357 & 39310 & 144 & 0.0000024\\\hline
\end{tabular}
\caption{Free energy and parameters for a polymer in a sea of blockers when
$\mathcal{V}>\mathcal{V}_{1}$.}%
\end{table}

The transition occurs between $v=0.041$ and $0.042$. Again we could find
analytically almost the entire solution both for $v\ll x$ and at the
transition. The solution is given in the Appendix. There we show that the
transition occurs at $v=0.04142$. We observe that as for the case of a
Gaussian random potential the ratio $w/m$ changes from $\gg1$ to $\ll1$ as $v$
approaches the transition from below. The values of $Y$ are seen to be
consistent with the assumption $Y^{3}\gtrsim\mathcal{V}_{1}$ . We also checked
that the free energy from Table II is lower than what one would obtain by
constraining $Y$ to be in Region I, i.e. $Y^{3}<\mathcal{V}_{1}$.

Finally for $\mathcal{V>V}_{2}$, where $\mathcal{V}_{2}\simeq\exp\left(
x^{2/(d+2)}L^{d/(d+2)}\right)  $ (Region III), we expect the behavior of the
chain to stay the same. This is because for jumps within a volume
$\mathcal{V}_{2}$ the situation reverts to the previously discussed scenario,
since the effective volume of interest which determines the statistics of the
free spaces is of order $Y^{3}$ and we don't expect $Y$ to be that large.

An important point to note is that if one performs a simulation with strict
self-avoiding-walks on a diluted lattice (with $x<1)$, one has $v\sim1$, and
hence one will always be in the delocalized phase and will not see any
localization effects \cite{barat}.

A few words are in order about the case of an annealed potential. This case
has been already investigated in the literature \cite{thirumalai}, and we will
review it briefly. The free energy in the annealed case reads (for $d=3$)%
\begin{equation}
F(R)\approx\frac{L}{R^{2}}+xR^{3}+\frac{vL^{2}}{R^{3}}.
\end{equation}
The second term represents the entropy cost of a fluctuation in the density of
obstacles that creates an appropriate spherical region of diameter $\sim R$.
It was assumed that the chain occupies a spherical volume, or at least
deviations from a spherical shape are not large \cite{thirumalai}. In the case
$v=0$ one obtains by minimizing the free energy that $R\sim(L/x)^{1/5}$, a
well known result. For $v>0$ the first term is irrelevant (and so is the
``stretching term'' of the form $R^{2}/L$) and one finds that $R\sim
(v/x)^{1/3}L^{1/3}$. In $d$ dimension, the size scales like $L^{1/d}$ when
$v>0$, which is larger than the $L^{1/(d+2)}$ dependence in the $v=0$ case.
There is no indication for a phase transition in these arguments, although some
authors \cite{thirumalai} speculate that it breaks down for large $v$ and a
transition to a Flory $L^{3/(d+2)}$ dependence takes place.

\section{Conclusions}

In this paper we have considered the case of a long molecule subject to a
combination of a random environment and a self-avoiding interaction. We have
found that in the limit of a very long chain and when the self-avoiding
interaction is weak, the conformation of the chain consists of many blobs with
connecting segments. The blobs are situated in regions of low potential for a
Gaussian random potential or in regions of low density of obstacles in the
random obstacles case. We have found that as the strength of the self-avoiding
interaction is increased the chain undergoes a localization-delocalization
transition since the binding energy per monomer is no longer positive. The
chain is then no longer attached to a particular location in the medium but
can easily wander around due to any perturbation. For a localized chain we
have estimated quantitatively the expected number of monomers in the blobs and
in the connecting segments.

\section{Acknowledgements}

YYG acknowledges support of the US Department of Energy (DOE), grant No. DE-FG02-98ER45686.

%\section{Appendix}
\appendix 

\section*{Appendix}

In this Appendix we solve the minimization conditions for the free energy,
both for $v\ll g$ and at or just bellow the transition. The equations obtained
by minimizing $f$ \ from Eq. (\ref{fepm}) with respect to $Y,w$ and $m$ are in
that order%
\begin{gather}
\frac{2Y}{m}-\frac{2m}{Y^{3}}-\frac{3vm^{2}}{Y^{4}}-\frac{6gwG}{Y}%
+\frac{9vgw^{2}G^{2}}{Y}=0,\label{miny}\\
-G^{2}+2wvG^{3}=f,\label{minw}\\
-\frac{Y^{2}}{m^{2}}+\frac{1}{Y^{2}}+\frac{2vm}{Y^{3}}=f, \label{minm}%
\end{gather}
where we used
\begin{equation}
G\equiv G(Y^{3})=3g\ln Y,
\end{equation}
and
\begin{equation}
f=\frac{1}{w+m}\left(  \frac{Y^{2}}{m}+\frac{m}{Y^{2}}+\frac{vm^{2}}{Y^{3}%
}-wG^{2}+vw^{2}G^{3}\right)  . \label{fa}%
\end{equation}

First, in the case $v\ll g$, some terms are very small and can be neglected
(compare with Table I). The equations become%
\begin{gather}
\frac{2Y}{m}-\frac{6gwG}{Y}+\frac{9vgw^{2}G^{2}}{Y}=0,\\
-G^{2}+2wvG^{3}=f,\\
-\frac{Y^{2}}{m^{2}}=f,\\
f=\frac{1}{w+m}\left(  \frac{Y^{2}}{m}-wG^{2}+vw^{2}G^{3}\right)  .
\end{gather}
These equations can now be solved analytically with the result given in
Eqs.\ (\ref{yvs},\ref{mvs},\ref{wvs}, \ref{fvs}).

At or very near the transition we put $f=0$ on the rhs of equations
(\ref{minm},\ref{minw}). Eq.\ (\ref{minw}) yields%
\[
w=\frac{1}{2vG}.
\]
The other two equations become%
\begin{gather}
-\frac{2m}{Y}\left(  -\frac{Y^{2}}{m^{2}}+\frac{1}{Y^{2}}+\frac{2vm}{Y^{3}%
}\right)  +\frac{vm^{2}}{Y^{4}}-\frac{3g}{4vY}=0,\\
-\frac{Y^{2}}{m^{2}}+\frac{1}{Y^{2}}+\frac{2vm}{Y^{3}}=0.
\end{gather}
Thus using the second of these equations to eliminate the first term in the
first, the first equation becomes%
\begin{equation}
Y^{3}=\kappa vm^{2}\equiv\kappa x^{6}/v^{3},
\end{equation}
where we defined $\kappa=4v/(3g)$ and $x^{3}=v^{2}m$. The second equation
becomes%
\begin{equation}
\kappa^{5/3}x^{2}-2x-\kappa^{1/3}=0,
\end{equation}
which has the solution%
\begin{equation}
x=\frac{1+\sqrt{1+\kappa^{2}}}{\kappa^{5/3}}.
\end{equation}
The values of $m$ and $Y$ follows from the relations $m=x^{3}/v^{2}$ and
$Y=\kappa^{1/3}x^{2}/v$, which coincide with equations (\ref{ytr}%
,\ref{mtr},\ref{wtr}). Eq.(\ref{feqz}) just follows from Eq.(\ref{fa}) and the
condition $f=0$.

For the case of random obstacles and $\mathcal{V\gtrsim}\exp(1/\sqrt{x})$ we
could also solve the minimization equations of the free energy both for $v\ll
x$ and at the transition. In the first case the equation reads%
\begin{gather}
\frac{2Y}{m}-\frac{2wG_{o}^{2}}{3Y\ln Y}-\frac{vw^{2}G_{o}^{3}}{Y\ln Y}=0,\\
-x+G_{o}^{2}+2wvG_{o}^{3}=f,\\
-\frac{Y^{2}}{m^{2}}=f,\\
f=\frac{1}{w+m}\left(  \frac{Y^{2}}{m}-xw+wG_{o}^{2}+vw^{2}Go^{3}\right)  .
\end{gather}
where
\begin{equation}
G_{o}=\left(  \frac{x}{3\ln Y}\right)  ^{1/3}.
\end{equation}
We display the solution in the form valid when $v$ is small enough so that
$\ln Y\gg1$. It reads%
\begin{align}
y  &  =\frac{1}{v}\frac{2G_{o}}{9(\ln Y)^{2}}\frac{1}{\sqrt{x-G_{o}^{2}}},\\
m  &  =\frac{1}{v}\frac{2G_{o}}{9(\ln Y)^{2}}\frac{1}{x-G_{o}^{2}},\\
w  &  =\frac{1}{v}\frac{2}{3G_{o}\ln Y},\\
f  &  =-x+G_{o}^{2}.
\end{align}
Since $G_{o}$ still contains $\ln Y$, the first equation has to be solved
numerically for $Y$ and then the rest of the solution follows.

At the transition the minimization equations read%
\begin{gather}
\frac{2y}{m}-\frac{2m}{Y^{3}}-\frac{3vm^{2}}{Y^{4}}+2wG_{o}\frac{d}{dy}%
G_{o}+3vw^{2}G_{o}^{2}\frac{d}{dy}G_{o}=0,\\
-\frac{Y^{2}}{m^{2}}+\frac{1}{Y^{2}}+\frac{2vm}{Y^{3}}=0,\\
-x+G_{o}^{2}+2wvG_{o}^{3}=0,
\end{gather}
with
\begin{equation}
\frac{d}{dy}G_{o}=-\frac{G_{o}^{4}}{xY}.
\end{equation}
Defining%
\begin{equation}
\sigma=\frac{4xv}{(3x+G_{o}^{2})(x-G_{o}^{2})},
\end{equation}
the solution reads%
\begin{align}
Y  &  =\frac{1}{v\sigma}\left(  \frac{1+\sqrt{1+\sigma^{2}}}{\sigma}\right)
^{2},\\
m  &  =\frac{1}{v^{2}\sigma^{2}}\left(  \frac{1+\sqrt{1+\sigma^{2}}}{\sigma
}\right)  ^{3},\\
w  &  =\frac{x-G_{o}^{2}}{2vG_{o}^{3}}.
\end{align}
The transition point ($v_{c}$) is determined from the equation%
\begin{equation}
\frac{Y^{2}}{m}+\frac{m}{Y^{2}}+\frac{vm^{2}}{Y^{3}}-w(x-G_{o}^{2}%
)+vw^{2}G_{o}^{3}=0,
\end{equation}
which becomes%
\begin{equation}
\frac{1+\sqrt{1+\sigma^{2}}}{\sigma}+\frac{\sigma}{1+\sqrt{1+\sigma^{2}}%
}+\frac{1}{\sigma}-\frac{(x-G_{o}^{2})^{2}}{4vG_{o}^{3}}=0.
\end{equation}
Actually since $G_{o}$ and $\sigma$ still depend on $\ln Y$ one has to solve
numerically two coupled nonlinear equations . The solution gives
$v_{c}=0.04142,\ y=302.8,\ m=30145,\ w=140$.

%\section{References}

\end{document}